\documentclass[jap,graphicx, reprint,unsortedaddress]{revtex4-1} % for checking your page length
\usepackage{amsfonts, amsmath, graphicx, feynmf}
\usepackage[version=3]{mhchem}

\begin{document}

\title{Spin accumulation in disordered topological insulator thin films} 
\author{Zhuo Bin Siu}
\affiliation{Computational Nanoelectronics and Nanodevices Laboratory, Electrical and Computer Engineering Department, National University of Singapore, Singapore} 
\author{Cong Son Ho}
\affiliation{Computational Nanoelectronics and Nanodevices Laboratory, Electrical and Computer Engineering Department, National University of Singapore, Singapore} 
\author{Mansoor B. A. Jalil} 
\affiliation{Computational Nanoelectronics and Nanodevices Laboratory, Electrical and Computer Engineering Department, National University of Singapore, Singapore} 
\author{Seng Ghee Tan} 
\affiliation{Data Storage Institute, Agency for Science, Technology and Research (A*STAR), Singapore} 

\begin{abstract}
Topological insulator (TI) thin films differ from the more commonly studied semi-infinite bulk TIs in that the former possesses both top and bottom surfaces where the surface states localized at different surfaces can couple to one another due to the finite thickness of the film. In the presence of an in-plane mangnetization TI thin films display two distinct phases depending on which of the inter-surface coupling or the magnetization is stronger. In this work, we consider a TI thin film system with an in-plane magnetization and calculate numerically the resulting spin accumulation on both surfaces of the film due to an in-plane electric field to linear order. We describe a numerical scheme for performing the Kubo calculation calculation in which we include impurity scattering and vertex corrections.  We find that the sums of the spin accumulation over the two surfaces in the in-plane direction perpendicular to the magnetization, and in the out of plane direction, are antisymmetric in Fermi energy about the charge neutrality point and are non-vanishing only when the symmetry between the top and bottom TI surfaces is broken. The impurity scattering, in general, diminishes the magnitude of the spin accumulation and can also change the sign of the spin accumulation at some Fermi energies where the accumulation is small.   
\end{abstract} 

\maketitle

\section{Introduction} 
Topological insulators (TIs) \cite{RMP82_3045, RMP83_1057, JPSJ82_102001} are an emerging class of materials possessing surface states with unique properties \cite{PRB78_195424,PRL102_146805, Sci329_61, PRL100_096407}. For example, the higher mobilities due to  the suppression of back scattering \cite{PRL109_066803, Nat460_1106} and the spin-momentum locking in the TI surface states make TI attractive candidates in prospective spintronics device applications \cite{NatPhy5_378,NatMat11_409}. 

In particular, the  spin-momentum locking leads to the inverse spin-galvanic effect \cite{PRL104_146802, PRB81_241410} where the passage of an in-plane electric field through a TI leads to a resulting spin accumulation. This spin accumulation in turn exerts a torque on the magnetization of an adjoining ferromagnetic (FM) layer or FM dopants which may be used to switch the magnetization direction for magnetic memory \cite{APE4_094201,JAP117_17C739} applications. Besides the more traditional picture that the spin accumulation exerts a torque via exchange coupling on the magnetization by acting like an effective magnetic field, it has also been proposed recently that the alignment of the spin on the FM side of the FM-TI interface on a FM-TI heterostructure to the spin accumulation direction on the TI side of the interface and the subsequent diffusion of the spin across the thickness of the FM layer may also contribute to the spin torque \cite{PRB93_125303}. Spin torques in magnetized TI systems have recently been measured experimentally, both in heterostructures with a FM layer deposited on top of the TI \cite{Nat511_449, PRL114_257202} as well as in magnetically doped TIs \cite{NatMat13_699}.

Spin accumulation in  magnetized \textit{semi-infinite} TI slabs due to in-plane direct current electric fields have previously been studied theoretically first in the clean limit \cite{PRL104_146802}, subsequently in the presence of disorder scattering \cite{PRB81_241410}, and most recently with the inclusion of vertex corrections \cite{PRB89_165307}. The spin accumulation in TI thin films has however not received much attention to date \cite{JPD49_135003} .  TI thin films \cite{PRB80_205401,PRB81_041307,PRB81_115407} differ from the more commonly studied semi-infinite slabs in that they possess both a top as well as a bottom surface where the surface states localized on both surfaces can couple to one another due to the finite thickness. From the device point of view, thinner TI films offer the advantage that their larger surface area to volume ratios enhance the contribution of the more practically useful surface states relative to the bulk states. From a theoretical point of view the simultaneous presence of the inter-surface coupling and in-plane magnetization leads to two distinct topological phases depending on whether the inter-surface coupling is stronger than the in-plane magnetization \cite{PRB83_195413}. 

We have earlier studied the spin accumulation in TI thin films subjected to an in-plane magnetization due to an in-plane electric field in the absence of disorder \cite{Ar1606_03812}.  (In contrast, the works on electric field induced spin accumulation in semi-infinite bulk TIs have focused on out-of-plane magnetizations and treated the in-plane magnetization perturbatively. ) Here we extend our previous study to include the effects of disorder scattering and vertex corrections, and use more realistic material parameters.  

This paper is organized as follow. We first introduce the model for the TI thin film system in the next section. We then describe the numerical evaluation of the Kubo formula for the spin accumulation. We then present and discuss our numerical results, and conclude with a brief summary of our findings. 

\section{Model} 
The Hamiltonian for a disordered TI thin film can be written as 
\begin{equation} 
	H = v (\vec{k}\times\hat{z})\cdot\sigma \tau_z + \lambda \tau_x + M_x\sigma_x + E_z\tau_z +  V_{\mathrm{imp}} \label{Ham0}.
\end{equation}
The $\vec{\sigma}$s in Eq. \ref{Ham0} refer to the real spin, while $\tau_z$ represents whether the states are localized nearer the top ($\langle \tau_z \rangle = +1$) or bottom ($\langle \tau_z \rangle = -1$) surface. The first term $v(\vec{k}\times\hat{z})\cdot\sigma \tau_z$ is hence the usual low energy Dirac fermion effective Hamiltonian for a TI with an additional factor of $\tau_z$ representing states localized near the top and bottom surfaces for the two signs of $\langle\tau_z\rangle$, and $\lambda \tau_x$ the inter-surface coupling. The $M_x\sigma_x$ term represents a fixed in-plane magnetization in the $x$ direction, and the $E_z\tau_z$ term an asymmetry between the top and bottom surfaces in the TI film. This asymmetry results from the top and bottom contact of the film being in contact with different materials -- in most experimentally grown TI thin film systems the bottom layer of the film is adjacent to the substrate whereas the top surface abuts either the vacuum or a ferromagnetic layer.  

We model the disorder scattering as non-magnetic Dirac delta point scatterers, 
\[
	V_{\mathrm{imp}} = \sum^{N_{\mathrm{imp}}}_{i} u\delta(\vec{r}-\vec{R}_i).
\]
and average over the impurity positions. 

The Hamiltonian Eq. \ref{Ham0} \textit{without} the $V_{\mathrm{imp}}$ term can be diagonalized analytically. However, the resulting analytic results are cumbersome and neither illuminating nor easy to work with. We therefore adopt a numerical approach in this work. For a given $\vec{k}$, Eq. \ref{Ham0} yields four values of eigenenergies $\epsilon(\vec{k},s)$ and corresponding eigenstates  $|\vec{k},s\rangle$ where $s=1,2,3,4$ distinguishes between the four eigenstates. Two of these eigenstates correspond to particle-like states in the dispersion relation, and the other two to hole-like states. The particle states and the hole states each consist of a pair of states with one member of the states localized closer to the top surface, and the other localized nearer the bottom surface \cite{Ar1606_03812}. These states are not degenerate for finite values of $M_x$. 

\section{Theory}
Our approach follows that of Ref. \onlinecite{PRB89_165307} loosely. We calculate the spin accumulation to linear order in the electric field using the Kubo formula. The non-equilibrium expectation value of an arbitrary observable $O$, $\langle O \rangle$ to linear order in a DC electric field in the $i$th direction, $\mathcal{E}_i$ is given by 
\begin{equation}
	\langle O \rangle = \lim_{\omega \rightarrow 0} -\frac{1}{\omega} \mathrm{Im}\Pi^R_{OJ_i}(\omega)E_i
\end{equation}
where $\Pi^R_{OJ_i}$ is the retarded correlation function between the observable $O$ and the current in the $i$th direction $J_i$, and $E_i$ is the electric field in the $i$th direction. $\Pi^R_{OJ_i}(\omega)$ can be evaluated using the Matsubara complex frequency summation formalism and is given by 
\begin{equation}
	\Pi^R_{OJ_i}(\omega) = -\frac{1}{\beta}\sum_{q_n} \mathrm{Tr} \mathcal{G}(iq_n+i\omega')O\mathcal{G}(iq_n)J_i \Big|_{i\omega' \rightarrow \omega} \label{PiOJ}
\end{equation}
where $\mathcal{G}$ denotes Matsubara Green's function, $iq_n$ is a Matsubara frequency, and the trace is taken over $\vec{k}$ and $s$.

\begin{figure}[ht!]
\centering
\includegraphics[scale=0.8]{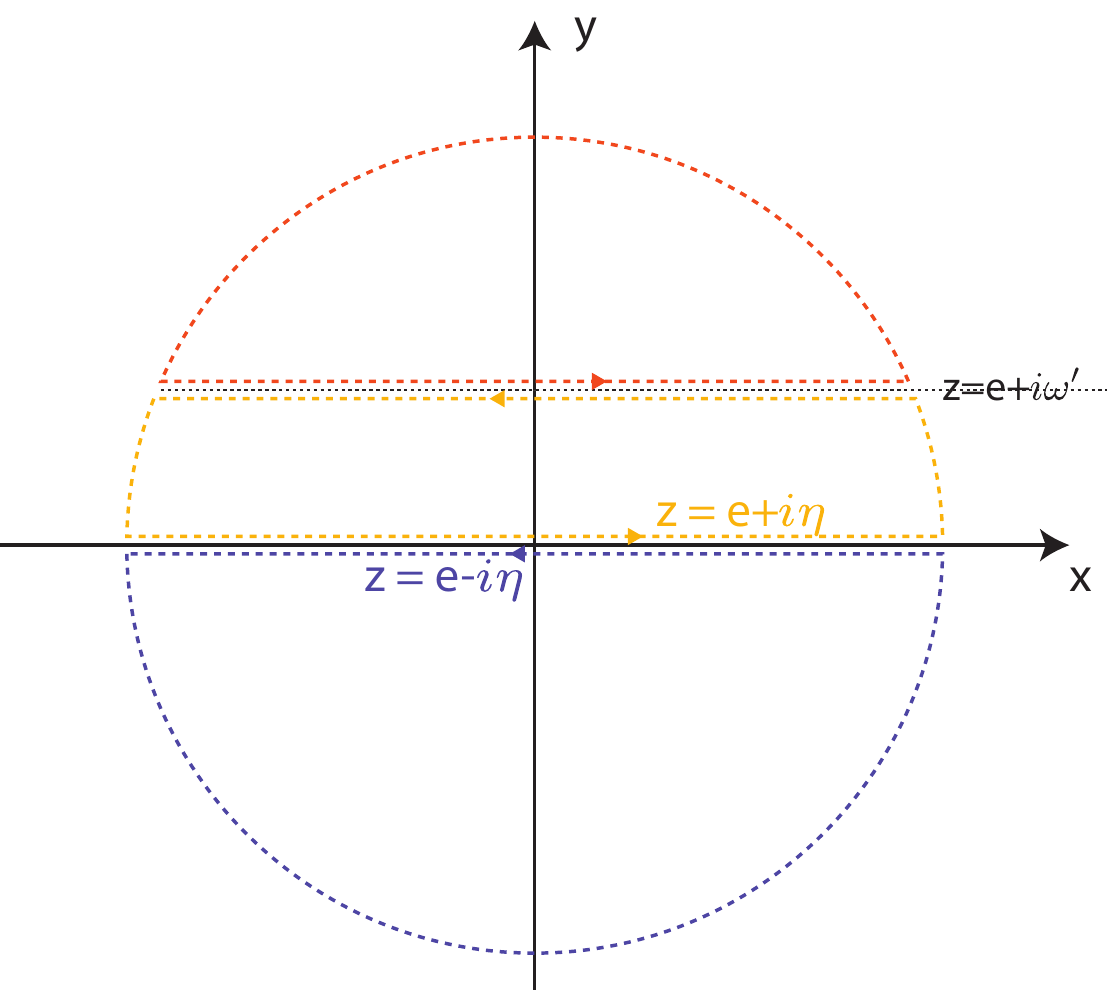}
\caption{  The integration contours for evaluating the Matsubara summation in the presence of impurity scattering. The vertical parts of the contours lie on both side of the two branch cuts of the integrand. } 
\label{gMatContour}
\end{figure}		    

Introducing
\begin{eqnarray} 
	f(z,z') &\equiv& \mathrm{Tr}\ \mathcal{G}(z)O\mathcal{G}(z')J_i  \label{fiqn} \\ 
	f^{CC'}(a, b)  &\equiv& \mathrm{Tr}\ G^C(a)OG^{C'}(b)J_i \label{fcc} 
\end{eqnarray} 
 where the $C,C' = A/R$  in Eq. \ref{fcc} for the \textit{a}dvanced / \textit{r}etarded Green's function 
 \begin{equation}
 G^{(R/A)}(e) \equiv (e- E_f - H - \Sigma^{R/A}) \label{Gra},
 \end{equation}
  the summation over $i q_n$ in Eq. \ref{PiOJ} can be converted to an integral over the usual Matsubara contour Fig. \ref{gMatContour}.  This  gives
\begin{eqnarray}
	\langle  O \rangle / E_i &=& \lim_{\omega \rightarrow 0} -\frac{1}{\omega} \mathrm{Im}\Pi^R_{OJ_i}(\omega) \nonumber \\
	&=& \lim_{\omega \rightarrow 0} \frac{1}{\omega} \left( \int_c \frac{i}{2\pi} \mathrm{Im} f(z + i\omega', z)n(z)  \right)\Big|_{\omega'\rightarrow \omega + i\eta} \nonumber \\ 
	&=& \lim_{\omega \rightarrow 0} \frac{1}{2\pi \omega} \left( \int_c \mathrm{Re} f(z + i\omega', z)n(z)  \right)\Big|_{\omega'\rightarrow \omega + i\eta}  \nonumber\\
	&=& \frac{1}{2\pi}  \int \mathrm{d}e\ \Big(  \lim_{\omega \rightarrow 0} \frac{n(e)-n(e+\omega)}{\omega} \nonumber\\
	&\ &\mathrm{Re} (f^{RR}(e+\omega, e) - f^RA)(e+\omega, e) \Big)  \nonumber \\
	&\approx& \frac{1}{2\pi} \mathrm{Re}( f^{RA}(0,0) - f^{RR}(0,0)) \nonumber \\ 
	&=& \frac{1}{2\pi} \mathrm{Re}  \mathrm{Tr} (  G^R(0)-G^A(0)) O G^R(0)J_i)  \label{TrGOGJ}
\end{eqnarray} 

We note that Eq. \ref{TrGOGJ} picks up both what has been termed the `interband' contributions \cite{Manchon} which give, in the clean limit, what is commonly called \textit{the} Kubo formula $\langle  O \rangle \propto \sum_{\alpha \neq \beta} \frac{n_\alpha-n_\beta}{(e_\alpha-e_\beta)^2} \mathrm{Im} (O_{\alpha\beta}J_{\beta\alpha})$ as well as the `intraband' contributions due to the electric field induced shift in the Fermi surface $\langle  O \rangle = \sum_{\vec{k}} \langle O(\vec{k})\rangle (\partial_e n) (\partial_{\vec{k}} e)\cdot\delta\vec{k}$. The interband term at a given Fermi energy is the sum of contributions over all the occupied states with energies below the Fermi energy, whereas the intraband contribution involves only states in the energy vicinity of the Fermi energy. 

In the first Born approximation, the retarded impurity self energy $\Sigma^R(e)$ appearing in the retarded Green's function Eq. \ref{Gra} is 
\begin{equation}
	\Sigma^R(e) = n_i u^2 \sum_{\vec{k},s=1-4} | \vec{k}, s \rangle \frac{1}{e - E_f - \epsilon(\vec{k},s) + i\eta} \langle \vec{k},s |. 
\end{equation}
where $n_i$ is the impurity concentration per unit area. 
Disregarding the real part of the self energy, we have 
\begin{eqnarray*}
	i \mathrm{Im} \Sigma^R(e) &=& n_i u^2 \sum_{\vec{k},s=1-4} | \vec{k}, s \rangle \mathrm{Im} \frac{1}{e - E_f - \epsilon(\vec{k},s) + i\eta} \langle \vec{k},s |  \\
	&=& n_iu^2 \sum_{\vec{k},s} |\vec{k}, s\rangle -i\pi\delta(e-E_f-\epsilon_{\vec{k},\sigma}) \langle \vec{k},s|. 
\end{eqnarray*}

\begin{figure}[ht!]
\centering
\includegraphics[scale=0.4]{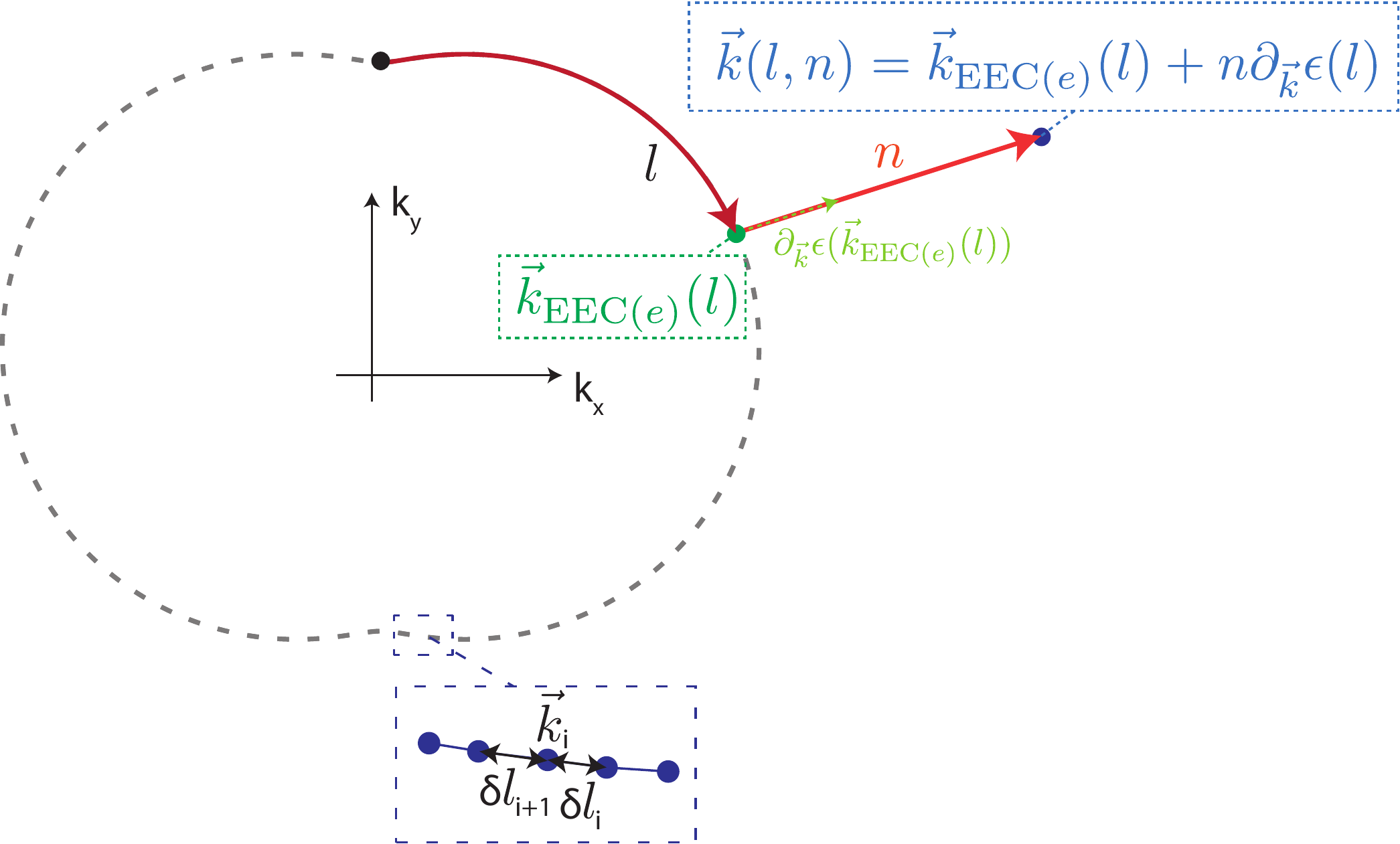}
\caption{  The gray dotted line represents the EEC for one of the $s$ bands at a given value of energy. The black dot at the top denotes the arbitrary origin for the path length $l$ on the EEC, the green dot the point on the EEC $\vec{k}_{\mathrm{EEC}(e)}(l)$, the dotted arrow next to it $\partial_{\vec{k}}  \epsilon(\vec{k}_{\mathrm{EEC}(e)}(l))$, and the bluish dot an arbitrary point in $k$ space labeled by the coordinates $(l,n)$.  The inset near the bottom shows that in an actual numerical calculation the EECs are constructed by algorithmically connecting discrete $\vec{k}$ points, represented by the solid purple circles, from which the spatial separation between adjacent points can be readily calculated.  } 
\label{gEECdiag}
\end{figure}

In order to evaluate the $\sum_{\vec{k}} f(\vec{k})\delta(e - \epsilon(\vec{k})) $ summation (we suppress the $s$ index here temporarily for notational simplicity ), we parametrize $k$ space as follows.  We denote a point on the EEC at energy $e$, and path distance $l$ from an arbitrary origin lying on the EEC as $\vec{k}_{\mathrm{EEC}(e)}(l)$. Since points on the EEC, by definition, have the same energy, the $k$-space vector normal to the EEC at a given $\vec{k}$ on the EEC is given by $\partial_{\vec{k}}  \epsilon(\vec{k}_{\mathrm{EEC}(e)}(l))$, which we subsequently write as $\partial_{\vec{k}} \epsilon(l)$ for short. An arbitrary point in $k$-space, not necessarily lying on the EEC (see Fig. \ref{gEECdiag}),  can then be parametrized by 
\[
\vec{k}(l, n) = \vec{k}_{\mathrm{EEC}(e)}(l) + n \partial_{\vec{k}}\epsilon(l).
\]

  This gives the infinitesimal  $k$-space area element $\mathrm{d}A = |\partial_{\vec{k}} \epsilon|\  \mathrm{d}l\ \mathrm{d}n$. This parametrization also gives (using the usual expansion for the Dirac delta of a function $\delta(f(x)) = \frac{\delta(x-x_0)}{|\partial_x f(x_0)|}$ ) 
\[
	\delta(e - \epsilon(\vec{k})) = \frac{ \delta(n) }{ |\partial_n \vec{k}(l, n=0)| } = \frac{ \delta(n) }{ |\partial_{\vec{k}} \epsilon(\vec{k})| }. 
\]

Putting everything together and using $\sum_{\vec{k}} \rightarrow \frac{1}{(2\pi)^2} \int \ \mathrm{d}\vec{k}$ gives
\begin{eqnarray*}
	\sum_{\vec{k}} f(\vec{k})\delta(e-\epsilon(\vec{k})) &=& \frac{1}{(2\pi)^2} \int\ \mathrm{d}{\vec{k}} f(\vec{k})\delta(e-\epsilon(\vec{k})) \\
	 &=& \frac{1}{(2\pi)^2} \int_{\mathrm{EEC}} \mathrm{d}l f(\vec{k})
\end{eqnarray*}
where the $\int_{\mathrm{EEC}}$ denotes integrating over the EEC. Now setting $e=E_f$, the retarded self energy is hence given by 
\begin{equation}
	\Sigma^R	\approx -i\frac{1}{4\pi} \sum_s \int_{\mathrm{EEC}(e=E_f,s)} \mathrm{d}l\ |\vec{k}, s\rangle\langle \vec{k}, s| \label{SReq} 
\end{equation}
where $\int_{\mathrm{EEC}(e=E_f,s)}$ means integrating along the $e=E_f$ EEC of the $s$th band. Although the parametrization in Fig. \ref{gEECdiag} may seem somewhat contrived, the end result Eq. \ref{SReq} turns out to be convenient to work with in a numerical calculation. For a given value of energy $e$ and, say, $k_x$, the Schroedinger equation $(H-e)|\vec{k},s\rangle$ can be cast into a generalized eigenvalue problem which can be numerically solved to give the eigenstates $|\vec{k},s\rangle$ and eigenvalues $k_y$ which satisfy the Schroedinger equation. Sweeping through the values of $k_y$ and solving the generalized eigenvalue problem at each value of $k_y$ hence gives a set of discrete $\vec{k}$ points that lie on the EECs.  Approximations to the EECs can then be constructed by linking up the discrete $k$ space points belonging to the same $s$ band. (See the inset of Fig. \ref{gEECdiag}. ) In our implementation we swept through both $k_y$ to obtain the corresponding $k_x$ points, as well through $k_x$ to obtain the corresponding $k_y$ points in order to increase the number of discrete $k$ space points for the construction of the EECs. The path distance between adjacent discrete points $\delta l$ on the EEC can be readily calculated from the points so that a numerical approximation to Eq. \ref{SReq} can be obtained as 
\[ 
	\Sigma^R	\approx -i\frac{1}{4\pi} \sum_i \sum_s  \ |\vec{k}_i, s\rangle\langle \vec{k}_i, s|  \frac{(\delta l_i + \delta l_{i+1})}{2}. 
\]

\begin{figure}[ht!]
\centering
\includegraphics[scale=0.6]{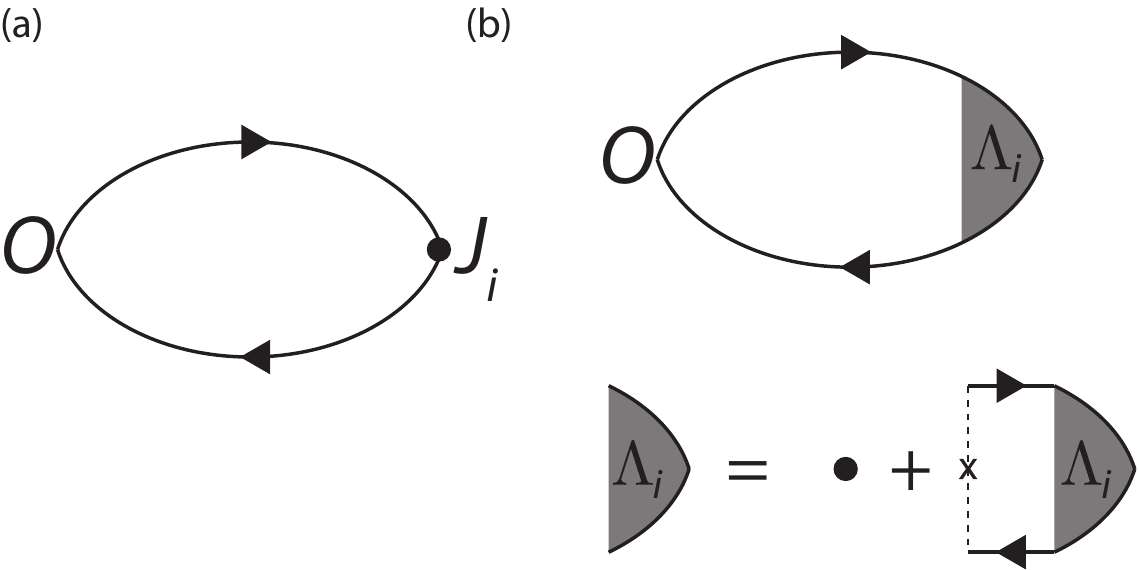}
\caption{  (a) The diagrammatic representation of Eq. \ref{TrGOGJ}, and (b) (top) the diagrammatic expression for $\langle \delta O \rangle$ incorporating the vertex correction $\Lambda$ and (bottom) the diagrammatic representation of the vertex correction  } 
\label{gFeynDiag}
\end{figure}

The expression for the non-equilibrium expectation value of  the observable $O$ Eq. \ref{TrGOGJ} can be expressed diagrammatically as Fig. \ref{gFeynDiag}(a). Whereas the expression already incorporates the effects of impurity scattering to all orders \textit{within} the upper and lower Green's function lines individually via the inclusion of the impurity scattering self energy within the Green's function, a more complete model would also incorporate scattering \textit{between} the upper and lower lines.  This can be accomplished by replacing the combination of $G^CJ_iG^{C'}$, ($C = R/A$) occurring in the lead-up to Eq. \ref{TrGOGJ}  with the vertex corrected version $G^C\Lambda^{CC'}G^{C'}$. Adopting the ladder approximation, $\Lambda$ is defined diagrammatically in the bottom panel of Fig. \ref{gFeynDiag}(b).

The Bethe-Salpeter equation for the Fig. \ref{gFeynDiag}(b) reads  
\begin{eqnarray}
	\Lambda^{CC'}_i &=& J_i + n_iu^2 \sum_{\vec{k}} G^C(\vec{k'}) \Lambda G^{C'}(\vec{k}) \nonumber \\
	\Rightarrow \Lambda^{CC'}_{i,\alpha\beta}  &=& J_{i,\alpha\beta} +  n_iu^2 ( \sum_{\vec{k}} G^C_{\alpha\gamma}(\vec{k'})\Lambda_{\gamma\delta}G^{C'}_{\delta \beta}(\vec{k'}) )  \label{HB2}
\end{eqnarray}
where $\Lambda^{CC'}_{i,\alpha\beta} \equiv \langle \vec{k},\alpha|\Lambda|\beta,\vec{k}\rangle$ ($\alpha$ and $\beta$ are the $s$ indices in the eigenstates $|\vec{k}, s\rangle$). Due to the absence of $\vec{J}$ on $\vec{k}$ here,  Eq. \ref{HB2} can be cast as a system of linear equations in the 16 $\Lambda_{i,\alpha\beta}$ matrix elements after the $\vec{k'}$ summation, which can then be solved to obtain all 16 matrix elements.

Incorporating the vertex  corrections, the final expression for $\langle \delta O \rangle$ becomes
\begin{equation}
	\langle \delta O \rangle / \mathcal{E}_i = \frac{1}{2\pi} \mathrm{Re} \mathrm{Tr} ( O G^R(0)( \Lambda^{RA}G^A(0) - \Lambda^{RR}G^R(0)) )  \label{TrOGLG}
\end{equation}
\section{Results and discussion} 
In the numerical results which follow, the parameters $v$ and $\lambda$ in the Hamiltonian Eq. \ref{Ham0} were obtained using the material parameters for \ce{Bi2Se3} from Ref. \onlinecite{PRB82_045122} via the approach of Ref. \onlinecite{AIPAdv6_055706}. The latter is in turn based on Ref. \onlinecite{PRB81_115407} .  Unless otherwise stated, we fix $E_z = 1\ \mathrm{meV} $ and $M_x = 50\ \mathrm{meV}$ in the results which follow.

\subsection{Bandstructure} 
In order to set the scene for the subsequent discussion, we first note that a quantum phase transition occurs in a thin film with in-plane magnetization and inter-layer coupling when $|M_x|=|\lambda|$ \cite{PRB83_195413}. We describe the bandstructure in the two regimes. 

Setting $E_z=0$ for now, let us first consider the limit where $\lambda = 0$. In this limit, the Hamiltonian reduces to $H = \big( v(\vec{k}\times\hat{z})\cdot\vec{\sigma} + M_x\sigma_x\tau_z \big) \tau_z$. It can be readily seen that $H$ is block diagonal in the $\tau$ degree of freedom with the two blocks representing the states localized nearer the top and the bottom surfaces of the thin film respectively, and that the Hamiltonian within each block takes the form of the Dirac fermion Hamiltonian  $\pm \big( (vp_y\pm M_x)\sigma_x - vp_x\sigma_y \big)$ which we are familiar from the more commonly studied semi-infinite TI slabs. The dispersion relation hence consists of two separate Dirac cones with the cone tips displaced from one another by $2M_x/v_f$ along the $k_y$ direction. 

\begin{figure}[ht!]
\centering
\includegraphics[scale=0.5]{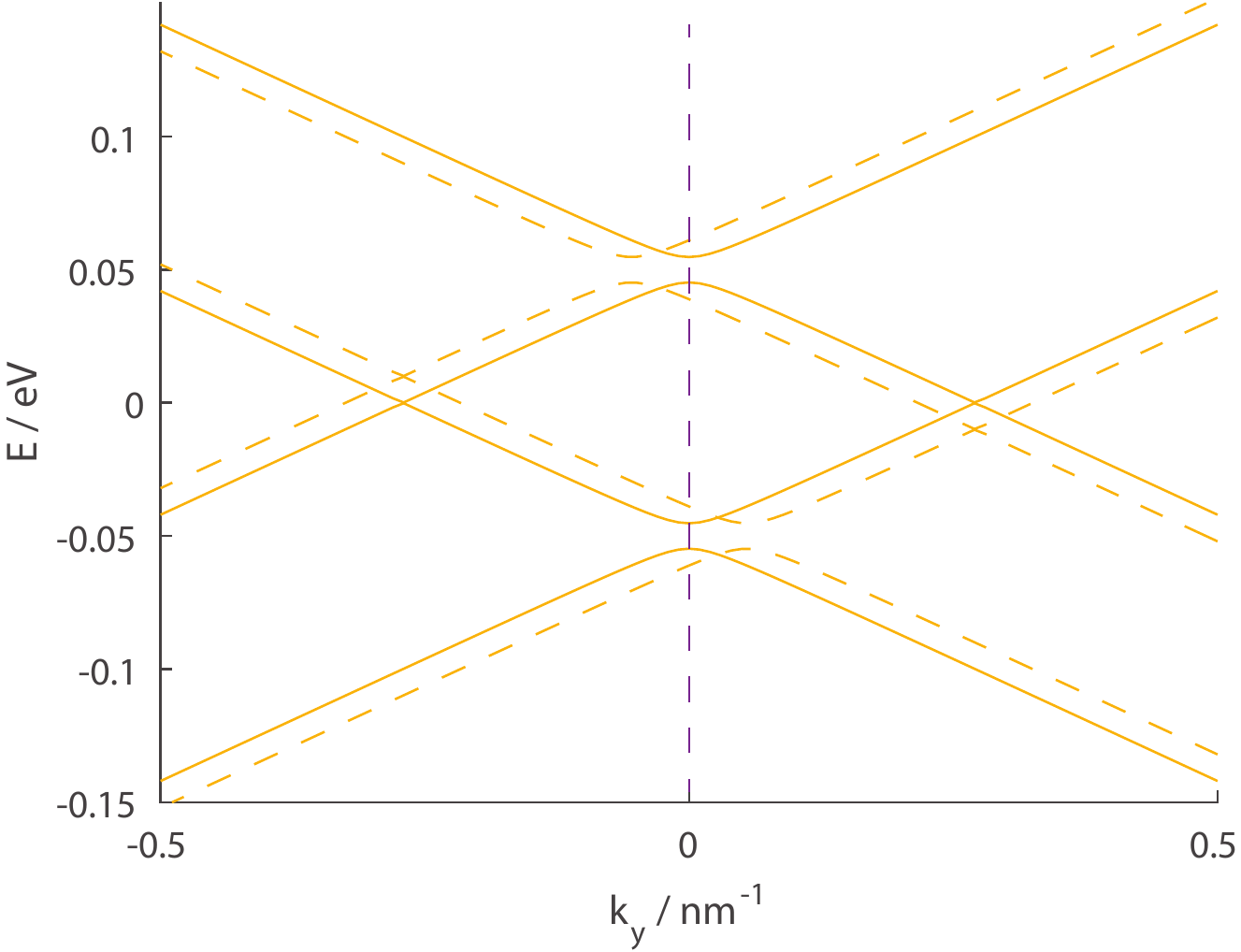}
\caption{  The dispersion relation of a $5\ \mathrm{nm}$ thick \ce{Bi2Se3} thin film at $k_x=0$ and $M_x = 50\ \mathrm{meV}$, at $E_z = 0$ (solid lines) and $E_z = 10\ \mathrm{meV}$ (dotted lines). The blue dotted line $k_y=0$ serves as a guide to the eye. } 
\label{gEw50} 
\end{figure}

A small finite value of $|\lambda| < |M_x|$ leads to an anti-crossing of the two Dirac cones where they cross each other in energy. We show in Fig. \ref{gEw50} the dispersion relation for a $5\ \mathrm{nm}$ film where $|M_x| = 50\ \mathrm{meV} > |\lambda| = 4.7\ \mathrm{meV}$. The cross sections of two distinct Dirac cones with Dirac points at $(k_x,k_y) = (0, \pm 0.26\ \mathrm{nm}^{-1})$ at the low energy $|E| < 20\ \mathrm{meV}$ regime and the avoided crossing of the two Dirac cones due to the inter-surface coupling at $k_y=0$ in the energy vicinity of $|E| \approx 50 \ \mathrm{meV}$ are clearly evident in the figure.   

\begin{figure}[ht!]
\centering
\includegraphics[scale=0.5]{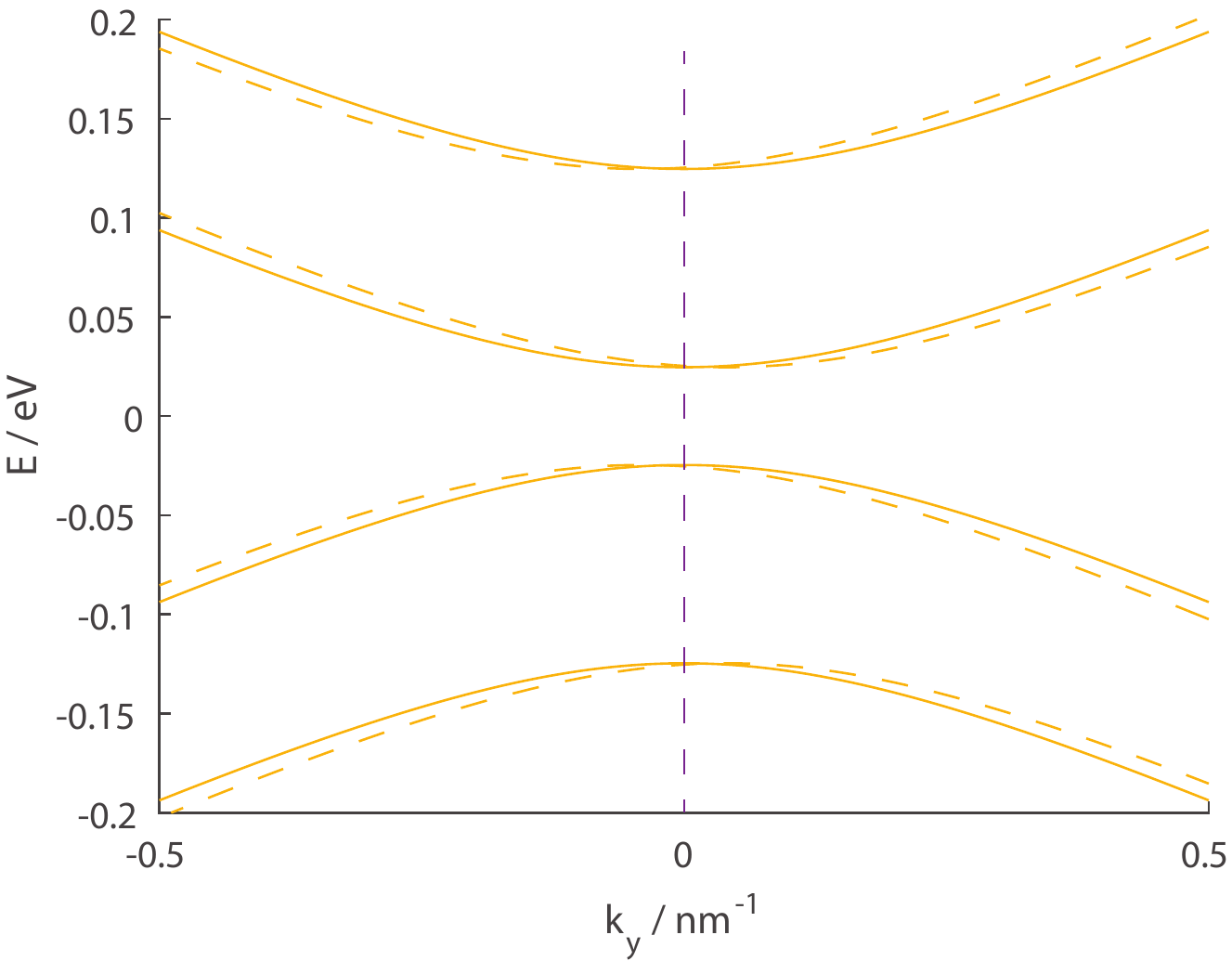}
\caption{  The dispersion relation of a $3\ \mathrm{nm}$ thick \ce{Bi2Se3} thin film at $k_x=0$ and $M_x = 50 \ \mathrm{meV}$, at $E_z = 0$ (solid lines) and $E_z = 10\ \mathrm{meV}$ (dotted lines). The blue dotted line $k_y=0$ serves as a guide to the eye.   } 
\label{gEw30} 
\end{figure}

As the inter-surface coupling strength is gradually increased (for example by decreasing the thickness of the film) , the energy splitting at the avoided crossing at $\vec{k}=0$ increases as well so that the hole and particle-like bands at $\vec{k}=0$ eventually touch, and with further increase of $|\lambda|$, cross each other and a band inversion occurs. Fig. \ref{gEw30} shows the dispersion relation of a $3\ \mathrm{nm}$ thick film where $|\lambda| = 75\ \mathrm{meV} > |M_x| = 50\ \mathrm{meV}$. In the strong inter-surface coupling regime we can no longer identify two distinct Dirac cones localized around the top and bottom surfaces of the TI film respectively, and a band gap opens up at $\vec{k}=0$. 

We now consider the effects of turning on $E_z$. The $E_z\tau_z$ term can be considered as a surface-dependent potential which shifts the energy of the states localized nearer the top (bottom) surface up (down). This is clearly evident from Fig. \ref{gEw50} where the two Dirac cones corresponding to the two surfaces can be distinctively recognized, and the Dirac point at negative (positive) $k_y$ corresponding to states localized nearer the top (bottom) surface is pushed up (down) in energy.  (We have exaggerated the value of $E_z$ to ten times that used in our subsequent numerical results to make the effect of $E_z$ more evident in the plot. ) The energy shift is also evident from Fig. \ref{gEw30} where the lower energy band at negative (positive) $k_y$ localized nearer the top (bottom) surface \cite{Ar1606_03812} is pushed up (down) in energy as well. 

\subsection{Spin accumulation} 
	We now proceed to calculate the spin accumulation due to an in-plane electric field using Eq. \ref{TrOGLG}. As we noted in the introduction, a key distinction between a TI thin film and a semi-infinite bulk TI is that the former has two surfaces. In experimental settings, the magnetization can either be supplied by a FM layer deposited on top of the TI thin film, or by magnetic doping. For the case of magnetic doping one might assume that the net spin torque experienced by the magnetization is due to the sum of the spin accumulations on both the top and bottom surfaces of the TI thin film. For a FM layer depposited on top of the film, it would be reasonable to assume that the top surface of the TI thin film will have a greater contribution to the spin torque than the bottom due to the closer proximity. We are however not aware of any studies that have been performed on how much more the contribution of the top surface is. We therefore calculate both the \textit{sum} of the spin accumulation on the top and bottom surfaces $\langle \sigma_i \rangle$, as well as their differences $\langle \sigma_i\tau_z \rangle$ in the directions perpendicular to the magnetization.  The spin accumulation on the top (bottom) surface is then given by  $\frac{1}{2} ( \langle \sigma_i \rangle +(-) \langle \sigma_i\tau_z \rangle)$. 

We focus here on the applied electric field parallel to the magnetization direction. An in-plane electric field in the $y$ direction \textit{perpendicular} to the magnetization direction does not give a significant $\langle \sigma_i \rangle$ and $\langle \sigma_i\tau_z \rangle$ spin accumulation in the $i=y,z$ directions perpendicular to the $x$ magnetization. While a $y$ electric field does lead to $\langle \sigma_x \rangle$ and $\langle \sigma_x\tau_z \rangle$ spin accumulation (not shown), neither of these exerts a torque on the $x$ magnetization.  

\begin{figure}[ht!]
\centering
\includegraphics[scale=0.3]{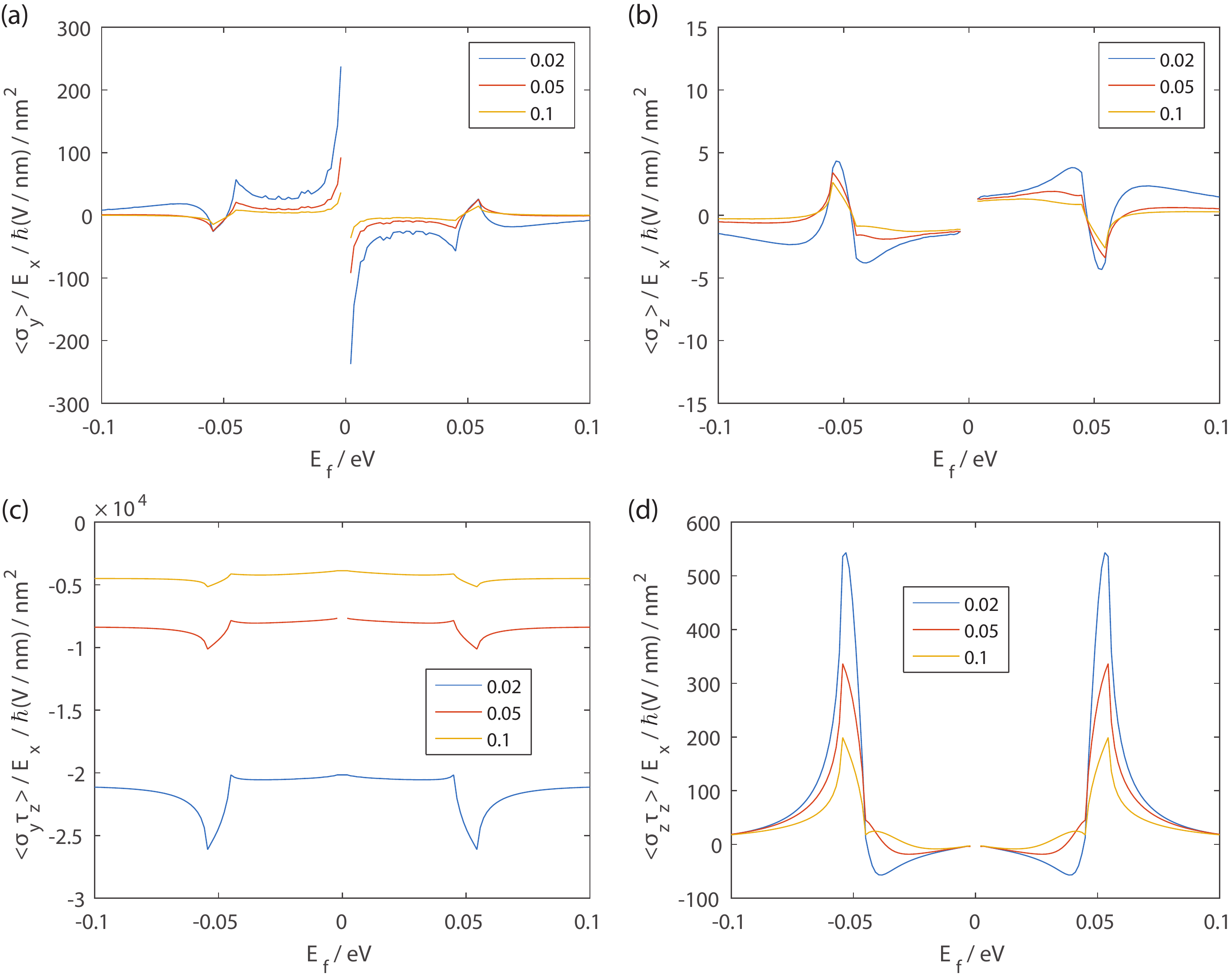}
\caption{  The sum and differences of the spin $y$ and $z$ accumulations due to an $x$ electric field in a $5\ \mathrm{nm}$ thick \ce{Bi2Se3} thin film with $E_z = 1\ \mathrm{meV}$ per unit electric field in the $x$ direction for three impurity scattering strengths $n_iu^2$ indicated in the panel legends in units of $\mathrm{eV^2 \AA^{-2}}$ } 
\label{gA1bComb} 
\end{figure}

Fig. \ref{gA1bComb} shows the spin $y$ and $z$ accumulations summed over all occupied states per unit $E_x$, the electric field in the $x$ direction parallel to the in-plane magnetization in the $|\lambda| < M_x$ regime with the same parameters as in Fig. \ref{gEw50} with the exception that here $E_z = 1\ \mathrm{meV}$.  There is no significant $\langle\sigma_x\rangle$ and $\langle\sigma_x\tau_x\rangle$ accumulation resulting from the in-plane $x$ electric field to the numerical precision of the $k$-space integration. Some of the data points at small $|E_f|$ are missing due to the numerical instabilities encountered. 

The $\langle\sigma_y\rangle$ and $\langle\sigma_y\tau_z\rangle$ accumulations are from intra-band contributions while the $\langle\sigma_z\rangle$ and $\langle\sigma_z\tau_z\rangle$ accumulations result from inter-band contributions. In particular, $-v$ times $\langle \sigma_y\tau_z\rangle$ gives the $x$ charge current whereas the $\langle \sigma_z \rangle$ spin accumulation can be attributed to an electric field induced effective magnetic field proportional to $\hat{\langle\sigma\rangle}\times\partial_{k_x}\hat{\langle\sigma\rangle}$ \cite{Ar1606_03812, Group, NatNano9_211}, $\hat{\langle\sigma\rangle}$ being the unit vector in the direction of the expectation value of the spin accumulation $\langle\vec{\sigma}\rangle$. Whereas the authors of the experimental study Ref. \onlinecite{PRL114_257202} had attributed the spin $z$ accumulation leading to their measured spin torques in a \ce{Bi2Se3}-FM heterostructure to the cubic momentum hexagonal warping, the electric field induced effective field may also have contributed to the spin accumulation.

  The magnitudes of the spin accumulations exhibit prominent peaks at $E_f = 54\ \mathrm{meV}$, and less prominent minor peaks at $E_f = 45\ \mathrm{meV}$. Comparing against Fig. \ref{gEw50}, we see that these peaks are related to key features in the dispersion relation -- the major peaks correspond to the energies where the higher energy band starts to emerge while the minor peaks correspond to the energy of the band top of the lower energy band at $\vec{k}=0$. 

\begin{figure}[ht!]
\centering
\includegraphics[scale=0.5]{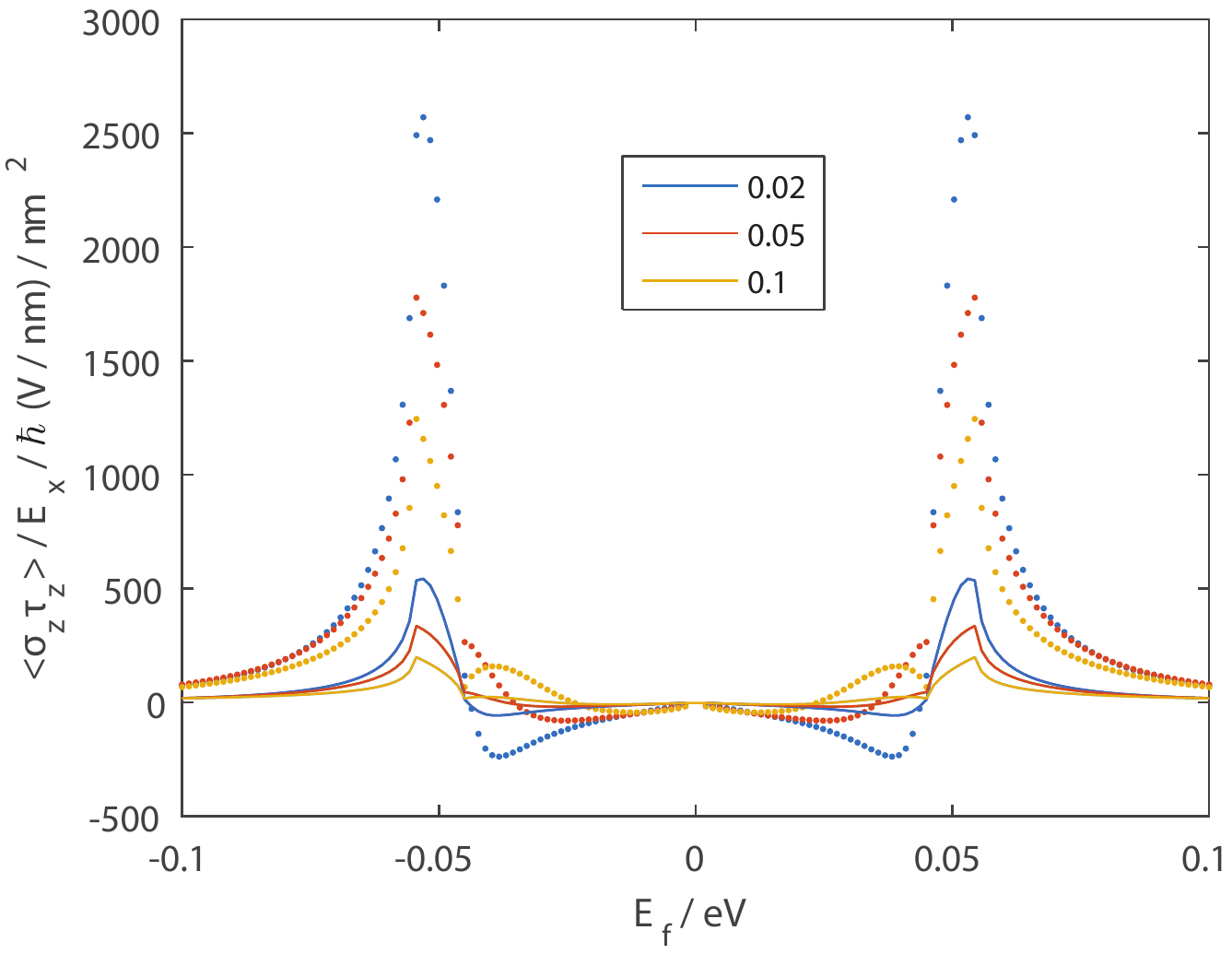}
\caption{$\langle \sigma_z\tau_z \rangle$ for a $5\ \mathrm{nm}$ thick \ce{Bi2Se3} thin film with $M_x = 50\ \mathrm{meV}$ and $E_z = 0\ \mathrm{meV}$ with (solid lines) and without (dotted lines) vertex corrections at various impurity scattering strengths $n_iu^2$ indicated in the legend in units of $\mathrm{eV^2 \AA^{-2}}$. }  
\label{gA1cVcComp} 
\end{figure}

The $\langle\sigma_i\rangle$s are antisymmetric with respect to $E_f$ whereas the $\langle \sigma_i\tau_z\rangle$s are symmetric.  In general, the magnitudes of the $\langle \sigma_i \rangle$s are orders of magnitude smaller than those of the $\langle \sigma_i \tau_z \rangle$s.  The relatively small value of the $\langle \sigma_i \rangle$s is due to their dependence on $E_z$, which has a small value here -- the $\langle\sigma_i\rangle$s vanish when $E_z = 0$ (not shown). The small value of $E_z$ does not lead to qualitative changes in the general trends for the $\langle\sigma_i\tau_z\rangle$s, as one can be seen from Fig. \ref{gA1cVcComp} where we plotted $\langle \sigma_z\tau_z \rangle$ for the same set of parameters as in Fig. \ref{gA1bComb} except that $E_z$ is set to 0 there. 

\begin{figure}[ht!]
\centering
\includegraphics[scale=0.5]{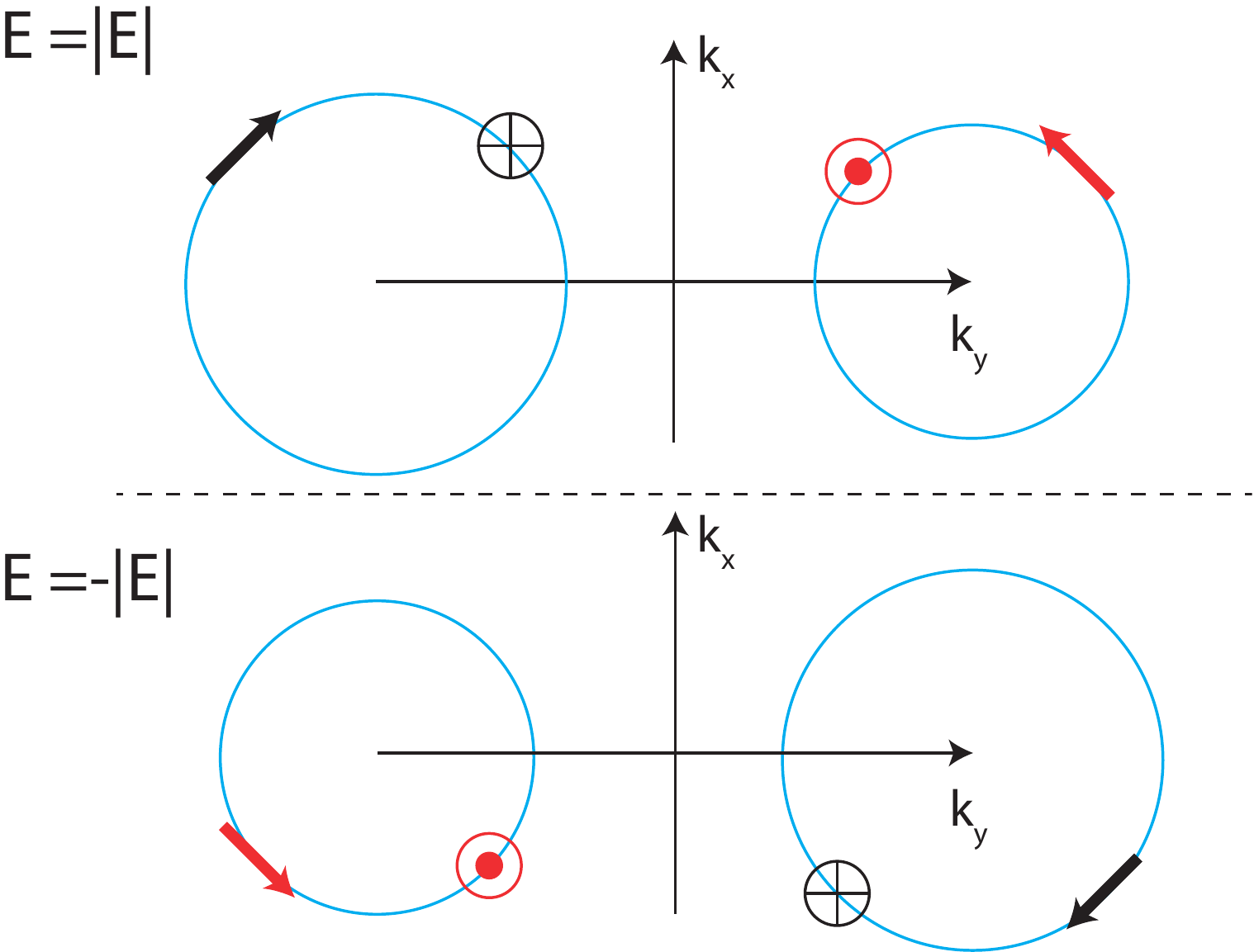}
\caption{A schematic plot of the relative signs of the spin accumulation directions on EEC points with a finite $E_z$ term showing how the asymmetry between the EECs for positive and negative $k_y$ are switched between $E > 0$ and $E < 0$. The spin accumulations in a given direction of the same color (black or red) have the same magnitudes; in the absence of the $E_z\tau_z$ term all the spin accumulations along a given direction at the points depicted would have the same magnitude. }  
\label{grelMag} 
\end{figure}

We explain why the $\langle \sigma_i \rangle$s are antisymmetric with respect to $E_f$. The spin accumulations plotted in Fig. \ref{gA1bComb} are essentially the spin accumulations integrated over the occupied states with every value of energy from negative infinity to the Fermi energy.  At any given value of energy,  the spin accumulation on the $k$ space points lying on the EEC at $E_z=0$ is antisymmetric in $\vec{k}$ space \cite{Ar1606_03812} and cancel out to zero after integrating over the EEC. A finite value of $E_z$ breaks this antisymmetry.  Notice from Figs. \ref{gEw50} and \ref{gEw30} that for a given value of energy magnitude $\epsilon$, the energy cross section EEC on $e=|\epsilon|$ is the reflection of the EEC for $w=-|\epsilon|$ about the $k_y = 0$ line, as indicated in Fig. \ref{grelMag}. 

Consider the symmetry properties of the Hamiltonian and the eigenstates. Including now the  $E_x x$ term (and omitting the impurity scattering term), the full Hamiltonian in the absence of $E_z$ now reads 
\[
	H_1 = v(k_y\sigma_x - k_x\sigma_y)\tau_z + \lambda \tau_x + M_x \sigma_x + E_x x.
\]
This Hamiltonian is invariant under the simultaneous transformation consisting of (i) spatial reflection about the $x$ axis, (ii) a $\pi$ rotation of the real spin along the spin $x$ direction, and (iii) a $\pi$ rotation of the $\tau$ degree of freedom along the $\tau_x$ direction which together lead to $k_y \rightarrow -k_y$, $\sigma_y \rightarrow -\sigma_y$, $\sigma_z \rightarrow -\sigma_z$,  $\tau_z \rightarrow -\tau_z$. These symmetries in turn imply that the spin $y$ and $z$ accumulations of states lying on a given $(k_x,k_y)$ on the EEC have the same magnitude but opposite signs as the accumulations of the the states on $(k_x,-k_y)$ 

The simultaneous transformations (i) $x\rightarrow -x$ (ii) $\pi$ spin rotation about the spin y direction and (iii) $\pi$ rotation about $\tau_y$ bring  $H_1 \rightarrow -H_1$. This implies that for a particle state for a given $(k_x,k_y)$ on the EEC the particle state with a positive energy $E$ has the same magnitude and sign of the spin y accumulation and opposite signs of the spin x and z accumulations for the hole state of energy $-E$ at $(-k_x,k_y)$.

If we now include the $E_z z$ term as well so that we have $H_2 = v(k_y\sigma_x - k_x\sigma_y)\tau_z + \lambda \tau_x + M_x \sigma_x + E_x x + E_z\tau_z$, we find the simultaneous transformations (i) spatial inversion on the $xy$ plane (ii) $\pi$ spin rotation about $\sigma_z$ and (iii) $\pi$ $\tau$ rotation about $\tau_y$ bring $H_2 \rightarrow -H_2$. This implies that for a particle state for a given $(k_x,k_y)$ on the EEC the particle state with an positive energy $E$ has the same magnitude but opposite signs of  spin x and y for the hole state of energy $-E$ at $-(k_x,k_y)$.  

Fig. \ref{grelMag} summarizes the relative signs of the spin accumulations on points related by symmetry on the EECs from the symmetry arguments above. It can be concluded from these symmetry properties that if the $E_z\tau_z$ term were absent so that the EECs are symmetrical about the $k_y=0$, the the spin y and z accumulations after summing over all the EEC $k$ space points would have canceled out by antisymmetry. The $E_z\tau_z$ term breaks the antisymmetry so that a finite spin accumulation remains after the EEC summation.  One can also conclude that even in the presence of a finite $E_z$,  the total spin y and z  accumulations for $E=\pm |E|$ after summing up over all the $k$-space points on the EEC at a given energy $E$,  $\langle \sigma_y \rangle (E)$ and $\langle \sigma_z \rangle (E)$, are \textit{symmetric} about $E = 0$. This symmetry in turn implies that the spin accumulation after integrating over all the occupied states, $\langle \sigma_i \rangle \equiv \int^{E_f}_{-\infty} \langle \sigma_i(E) \rangle \ \mathrm{d}E$ is \textit{anti}-symmetric. To see this, consider a $f(e)$ such that $\int^0_{-\infty} f(e)\ \mathrm{d}e = 0$ where $f(e)$ be symmetric in $e$. For a infinitesimal  $\delta e$ we then have $f(\delta e) = f(-\delta e)$ so that $\int^{\pm \delta e}_{-\infty} f(e)\ \mathrm{d}e = \pm f(\delta e)\delta e$. $\int^{E_f}_{-\infty} f(e)\ \mathrm{d}e$ is hence antisymmetric.

We now turn our attention back to other features of Fig. \ref{gEw50}. As one might expect, the magnitude of the spin accumulation in general decreases with increased impurity scattering characterized by $n_iu^2$. 

\begin{figure}[ht!]
\centering
\includegraphics[scale=0.5]{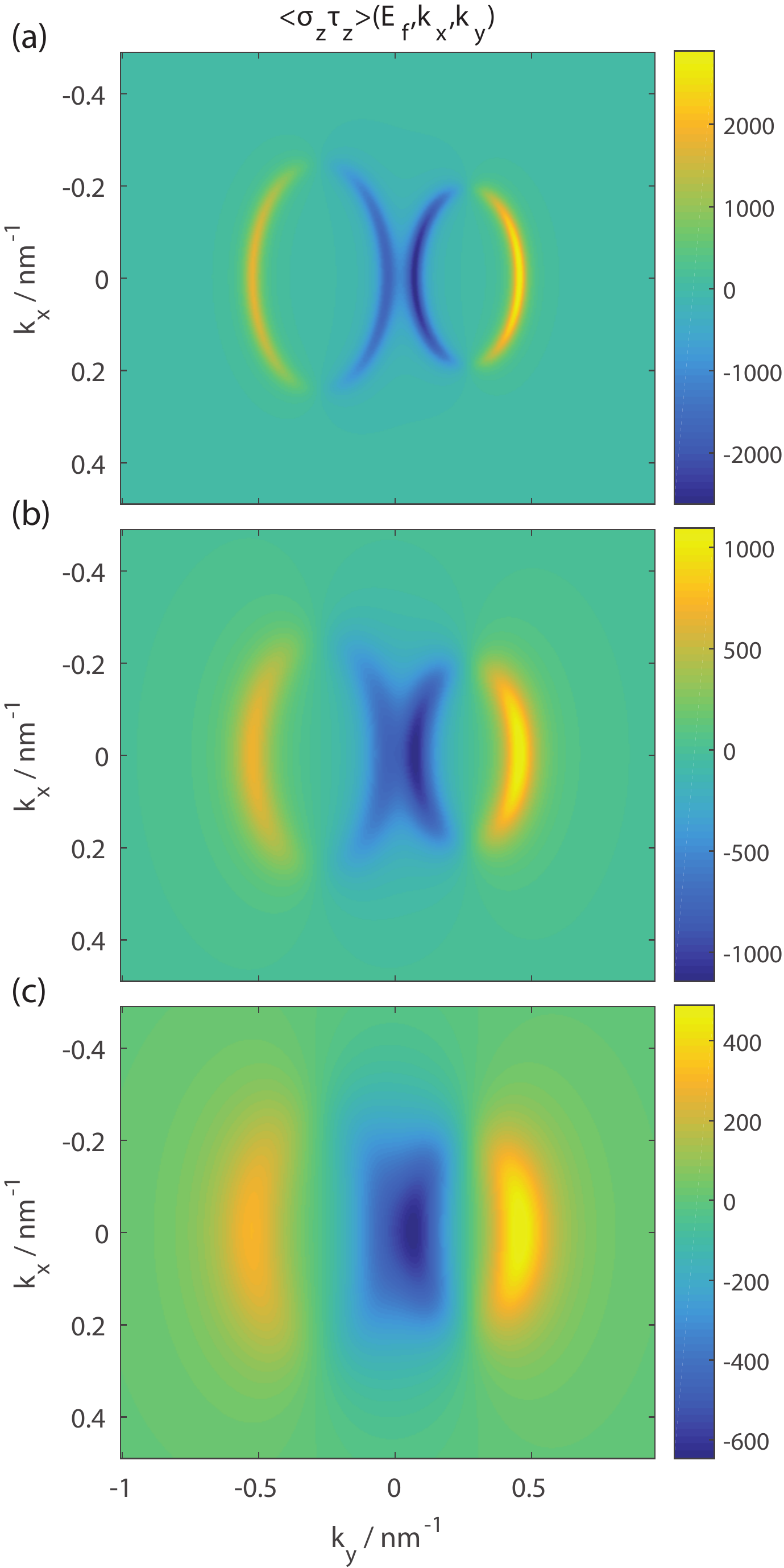}
\caption{ The quantity being summed over in $k$-space in the evaluation of Eq. \ref{TrOGLG} for the parameters in Fig. \ref{gEw50} at $E_f = 4\ \mathrm{meV}$ and (a) $n_iu^2 = 0.02\ (\mathrm{eV\AA^{-1}})^2$, (b) $n_iu^2 = 0.05\ (\mathrm{eV\AA^{-1}})^2$ and (c) $n_iu^2 = 0.1\ (\mathrm{eV\AA^{-1}})^2$. }
\label{gnoiseSmear} 
\end{figure}

While the impurity scattering does not change the symmetry properties of $\langle\sigma_i\rangle$ and $\langle \sigma_i\tau_z\rangle$  with respect to the sign of $E_f$ or the peaking of the spin accumulations at energies where near the higher energy bands emerge,  stronger impurity scattering does lead to small shifts in the  exact Fermi energies at which the magnitude of the spin accumulation peaks. The impurity scattering may also change the sign of the spin accumulations away from the peaks, as can been seen for the $\langle \sigma_y \rangle$ accumulation in the vicinity of $|E| > 50\ \mathrm{meV}$ and the $\sigma_z\tau_z$ accumulation. Fig. \ref{gnoiseSmear} shows the effects of increasing impurity scattering on the example of the $\langle \sigma_z\tau_z\rangle$ spin accumulation at small $|E_f|$.  

The impurity scattering leads to a smearing of the $k$-space features of the $k$-space integrand in Eq. \ref{TrOGLG}. This smearing out can perhaps be better understood by examining the equation for $\langle O \rangle$ in the absence of vertex corrections Eq. \ref{TrGOGJ}, in which $(G_\alpha^R(0)-G_\alpha^A(0))$ ($\alpha$ is a collective index for $(\vec{k},s)$ ) appears explicitly. The imaginary part of this term is approximately a Lorentzian distribution in $E_f$ centered around $E_f = E_\alpha$ with the energy width of the Lorentzian distribution characterized by the impurity scattering strength. The Lorentzian distribution reduces to a Dirac delta in the clean limit so that the plots in Fig. \ref{gnoiseSmear} will only pick up contributions in the $k$ space vicinity of the $E=E_f$ EECs. In the presence of finite scattering,  the $k$-space points in the vicinity of the these EECs will also pick up contributions from a broader neighborhood of the energy values in the vicinity of $E_f$. This however does not usually totally compensate for the energy broadening fall in the magnitudes of the contribution  $\vec{k}$ space points falling exactly on the $E=E_f$ EECs and so the spin accumulation magnitude typically drops with increased scattering.

We finally make a short comment on the effects of the vertex corrections. Fig. \ref{gA1cVcComp} shows an exemplary plot of the $\sigma_z\tau_z$ accumulation with the same parameter range as in Fig. \ref{gA1bComb} with (solid lines) and without (dotted lines) the vertex corrections calculated using Eqs. \ref{TrOGLG} and \ref{TrGOGJ} respectively. 
 
The inclusion of the vertex corrections does not change the main qualitative features of the spin accumulations mentioned earlier. The vertex corrections do, in general, lead to a reduction in the magnitude of the spin accumulation and a shift in the exact values of the Fermi energies where the spin accumulations switch sign. These comments also apply to the $\langle\sigma_y\rangle$,$\langle\sigma_z\rangle$ and $\langle\sigma_y\tau_z\rangle$ spin accumulations not shown here. 

\begin{figure}[ht!]
\centering
\includegraphics[scale=0.3]{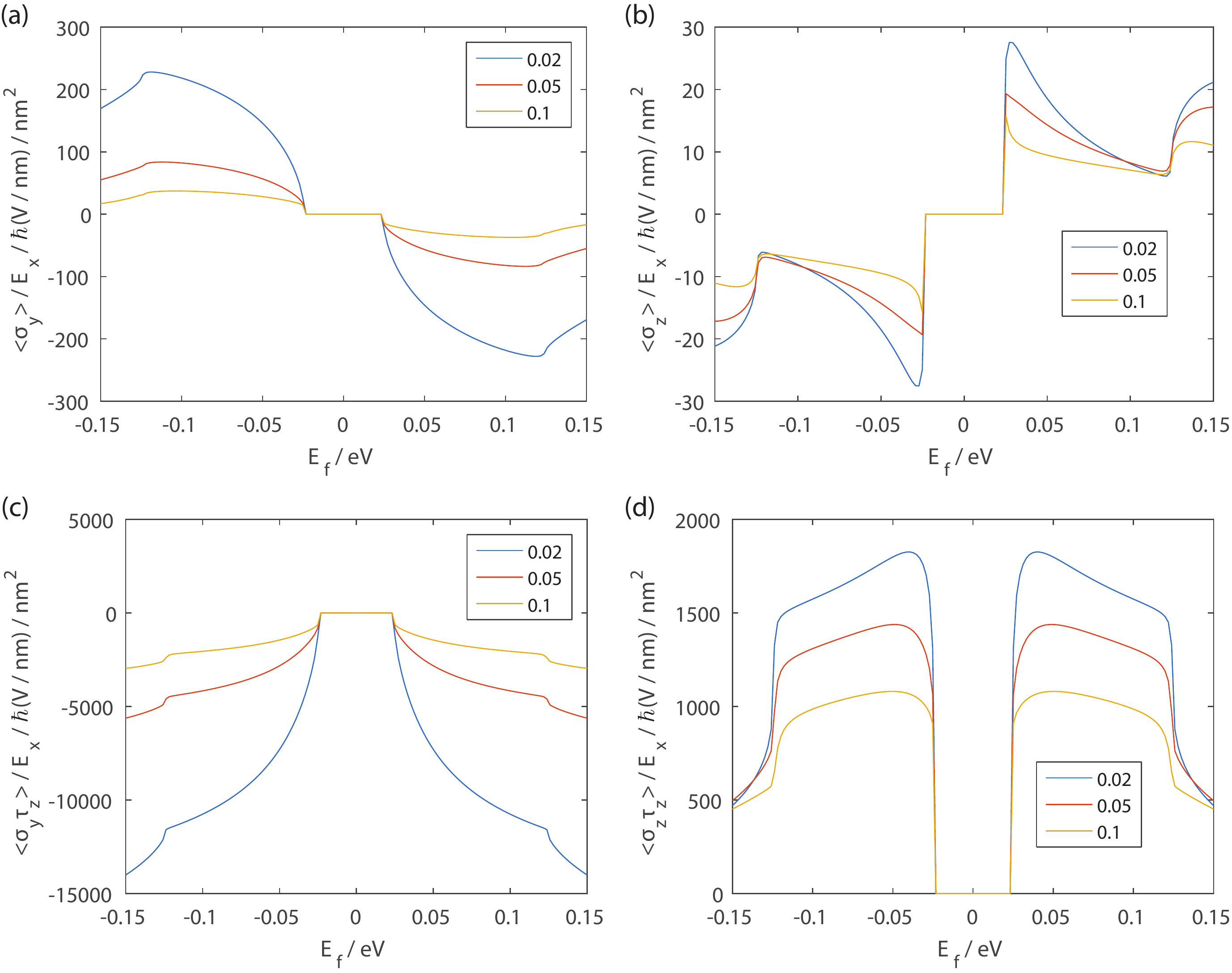}
\caption{  The sum and differences of the spin $y$ and $z$ accumulations due to an $x$ electric field in a $3 \mathrm{nm}$ thick \ce{Bi2Se3} thin film with $E_z = 1\ \mathrm{meV}$ per unit electric field in the $x$ direction for three impurity scattering strengths $n_iu^2$ indicated in the panel legends in units of $\mathrm{eV^2 \AA^{-2}}$ } 
\label{gA2bComb} 
\end{figure}

We now shift our attention to the spin accumulation in the strong inter-surface coupling regime. Fig. \ref{gA2bComb} shows the spin accumulations calculated for the $3.0\ \mathrm{nm}$ thick \ce{Bi2Se3} thin film for which we have plotted the dispersion relation in Fig. \ref{gEw50}.

The results in Fig. \ref{gA2bComb} share some qualitative similarities with those in Fig. \ref{gA1bComb} for the thicker film. $\langle \sigma_y \rangle$ and $\langle \sigma_z \rangle$ are antisymmetric in $E_f$, while $\langle \sigma_y\tau_z \rangle$ and $\langle \sigma_z\tau_z \rangle$ are antisymmetric. The magnitudes of the spin accumulations similarly decrease with impurity scattering. The major qualitative differences can be attributed to their differing bandstructures. In place of the peaks in the spin accumulation magnitudes occurring at two values of $|E_f|$ in the thicker film,  Fig. \ref{gA2bComb} has an energy range $|E| < 0.024\ \mathrm{eV}$ over which the spin accumulation is zero. This energy range corresponds to the inter-surface coupling induced bandgap (Fig. \ref{gEw30} ). Outside the bandgap, there is only one value of energy $|E| = 0.12\ \mathrm{eV}$ at which the spin accumulation is either maximal or minimal. This energy value corresponds to that at which the higher energy band emerges.  

\begin{figure}[ht!]
\centering
\includegraphics[scale=0.3]{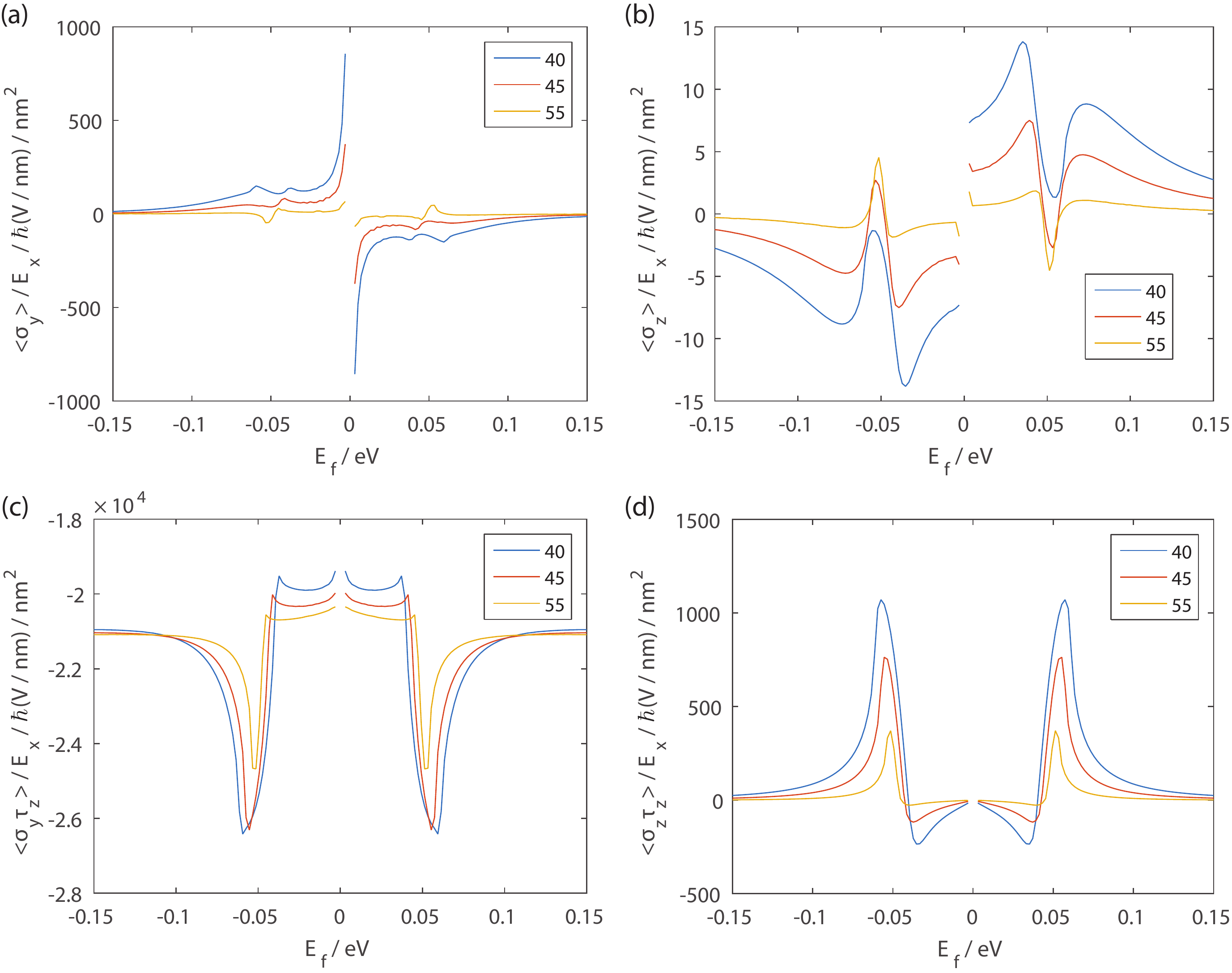}
\caption{  The spin accumulation for \ce{Bi2Se3} thin films of the thicknesses indicated in the legend (in \AA) at $n_iu^2 = 0.02\ \mathrm{eV\AA^{-1}}^2$, $E_z = 1\ \mathrm{meV}$ and $M_x =50\ \mathrm{meV}$ in the weak inter-layer coupling regime } 
\label{gA789} 
\end{figure}

\begin{figure}[ht!]
\centering
\includegraphics[scale=0.3]{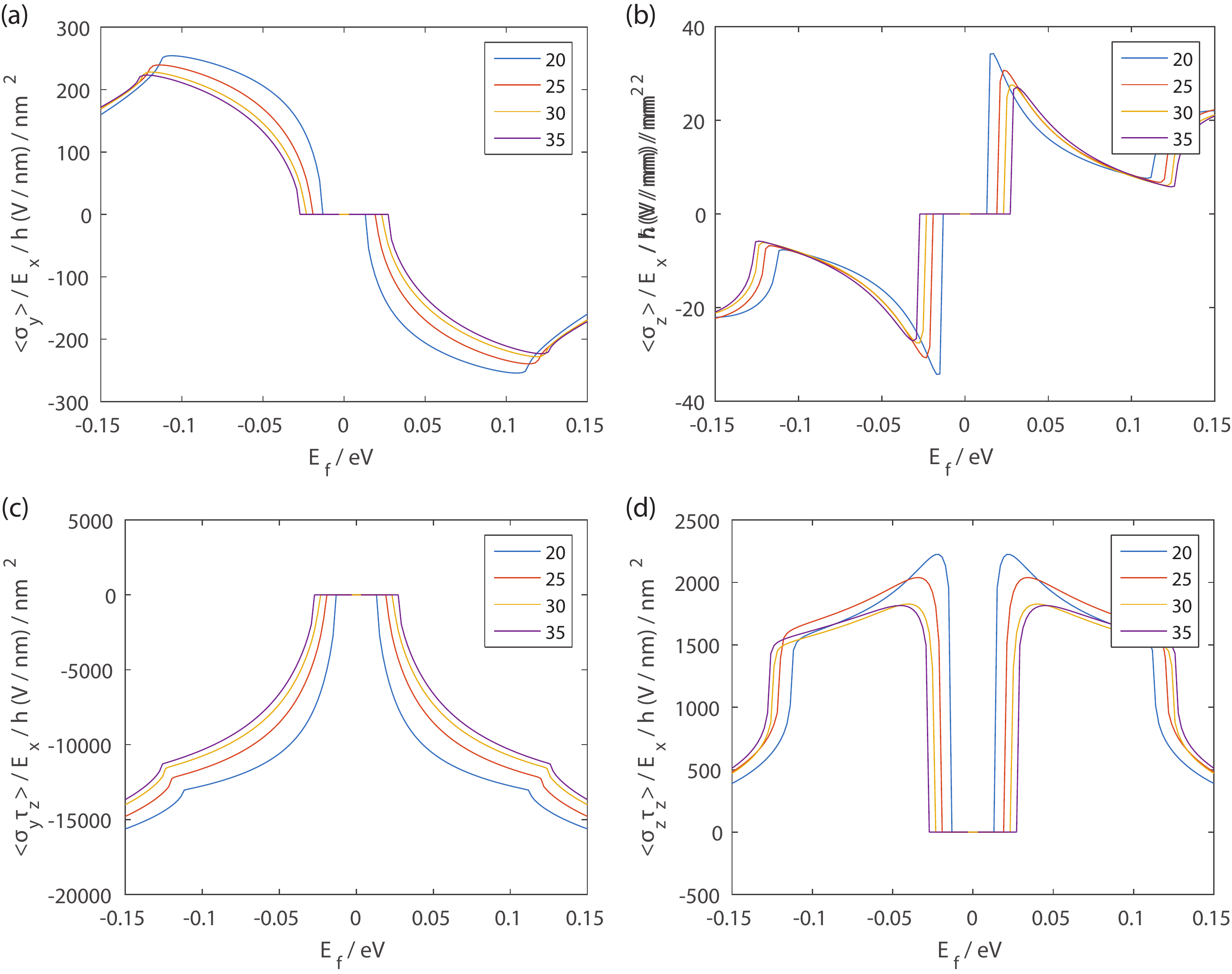}
\caption{  The spin accumulation for \ce{Bi2Se3} thin films of the thicknesses indicated in the legend (in \AA) at $n_iu^2 = 0.02\ \mathrm{eV\AA^{-1}}^2$, $E_z = 1\ \mathrm{meV}$ and $M_x =50\ \mathrm{meV}$ in the strong inter-layer coupling regime } 
\label{gA1011} 
\end{figure}

For completeness we plot in Figs. \ref{gA789} and \ref{gA1011} the spin accumulations in the weak and strong inter-layer coupling regimes respectively for various thicknesses of \ce{Bi2Se3} thin film with the same parameters as in the earlier figures. (As a reminder $M_x = 50\ \mathrm{meV}$ for all the figures but the interlayer coupling strength $\lambda$ does vary with the film thickness and is smaller (greater) in magnitude than $M_x$ for the smaller (larger) thicknesses depicted. ) 
 
The Fermi energy values at which the spin accumulations peak vary slightly with the thin film thicknesses. This is due to the dependence of the Fermi velocity $v_f$ and inter-layer coupling strength $\lambda$ appearing in the Hamiltonian Eq. \ref{Ham0} with the film thickness. In general, the maximum magnitude of the spin accumulations increase with decreasing thickness of the film. An interesting exception is for the $\langle \sigma_z \rangle$ and $\langle \sigma_y \rangle$  accumulations where the sign of the spin accumulation for the largest thickness depicted is reversed relative to that at smaller thicknesses.

\section{Conclusion} 
In this work we calculated the spin accumulation due to an in-plane electric field on the top and bottom surfaces of a TI thin film with an in-plane magnetization. We described the numerical scheme for this calculation incorporating disorder scattering in the first Born approximation and vertex corrections at the ladder level. We distinguished between the two cases where the in-plane magnetization is stronger, and where it is weaker, than the inter-surface coupling. In the first case, the states localized at the top and bottom surfaces give two distinct Dirac cones near the charge neutrality point whereas in the second case there are no longer two distinct cones and a bandgap opens up. We then showed that these bandstructure differences give rise to different qualitative trends in the spin accumulation as the Fermi energy is varied. In the weak inter-surface coupling regime the spin accumulation exhibits two distinct peaks as the Fermi energy is increased beyond the charge neutrality point corresponding to the band top of the lower energy band and the band bottom of the higher energy band at $\vec{k}=0$. In the strong inter-surface coupling regime the spin accumulation is zero when the Fermi energy falls inside the band gap and exhibits a kink when the energy is increased beyond the $\vec{k}=0$ band bottom of the higher energy band. 

We showed and explained why the $\langle \sigma_y \rangle$ and $\langle \sigma_z \rangle$ spin accumulations are antisymmetric with respect to the Fermi energy. These two quantities, which correspond to the sum of the spin accumulations on the top and bottom surfaces of the film, acquire finite values only when the symmetry between the top and bottom surfaces is broken so that the contributions from the two surfaces, which are of opposite signs, do not cancel out exactly. We showed that the impurity scattering and vertex corrections, in general, reduce the magnitude of the spin accumulation except at some values of Fermi energies where the magnitude is low in which case the sign of the spin accumulation may flip. The magnitude of the peak spin accumulation increases with decreasing thickness of the thin film. 

% The dependence of the sign and magnitude of the spin accumulation on the Fermi energy, which may also be affected by impurity doping, the asymmetry between the two surfaces of the film and the film thickness may account for the variation of the spin Hall angles experimentally measured by different groups.

\section{Acknowledgments} 
The authors acknowledge the Singapore National Research Foundation for support under NRF Award Nos. NRF-CRP9-2011-01 and NRF-CRP12-2013-01, and MOE under Grant No. R263000B10112.


\begin{thebibliography} {99}

% a collection of TI review papers 
	\bibitem{RMP82_3045} 	M. Z . Hasan and C. L. Kane,  Rev. Mod. Phys., \textbf{82}, 3045 (2010).
	\bibitem{RMP83_1057} X.-L. Qi ad S.-C. Zhang, Rev. Mod. Phys., \textbf{83}, 1057 (2011). 
	\bibitem{JPSJ82_102001} Y. Ando, J. Phys. Soc. Jpn. \textbf{82}, 102001 (2013). 

% oh haha laundry list of `exotic propties' 
	\bibitem{PRB78_195424} X.-L. Qi, T. L. Hughes and S.-C. Zhang, Phys. Rev. B \textbf{78}, 195424 (2008). % The scary shit Topologicial Field Theory paper which gets cited for all kinds of stuff 
	\bibitem{PRL102_146805}  A. M. Essin, J. E. Moore and D. Vanderblit, Phys. Rev. Lett. \textbf{102}, 146804 (2009).  % 	The originial `mangetoelectric polarizability and axsion electrody in crystalline insulators paper 
	\bibitem{Sci329_61} R. Yu \textit{et al.}, Science \textbf{329}, 61 (2010). % Quantized AHE in magnetic TIs
	\bibitem{PRL100_096407} L. Fu and C. L. Kane, Phys. Rev. Lett. \textbf{100}, 096407 (2008). % majorana fermions 


% dat backscatterinfg is suppressed in TIs
	\bibitem{PRL109_066803} A. A. Taskin \textit{et al.}, Phys. Rev. Lett. \textbf{109}, 066803 (2012). 
	\bibitem{Nat460_1106} P. Roushan \textit{et al.}, Nature \textbf{460}, 1106 (2009).

% Eh just go and cite for `spintronics applications' lah 
	\bibitem{NatPhy5_378} J. Moore, Nat. Phys. \textbf{5}, 378 (2009). 
	\bibitem{NatMat11_409} D. Pesin and A. H. MacDonald, Nature Mat. \textbf{11}, 409 (2012). 

% inverse spin-galvanic effect 
	\bibitem{PRL104_146802} I. Garate and M. Franz, Phys. Rev. Lett. \textbf{104}, 146802 (2010).  % no disorder 
	\bibitem{PRB81_241410} T. Kokoyama, J. Zang and N. Nagaosa, Phys. Rev. B \textbf{81}, 241410 (2010). % got disorder but no vertex corrections 
	
% lol our group's TI magnetic memory cell paper 
	\bibitem{APE4_094201} T. Fujita, M. B. A. Jalil and S. G. Tan, Appl. Phys. Exp. \textbf{4}, 094201 (2011). 
	\bibitem{JAP117_17C739} M. B. A. Jalil, S. G. Tan and Z. B. Siu, J. Appl. Phys. \textbf{117}, 17C739 (2015).

% the Manchon kahkia spin torque via spin diffusion paper 
	\bibitem{PRB93_125303} M. H. Fischer \textit{et al}, Phys. Rev. B \textbf{93}, 125303 (2016). 

% Exp spin torque papers
	% Eh ok these two are the FM on top of TI papers 	
		\bibitem{Nat511_449} A. R. Mellnik \textit{et al.}, Nature \textbf{511}, 449 (2014).  
		\bibitem{PRL114_257202} Y. Wang \textit{et al.}, Phys. Rev. Lett. \textbf{114}, 257202 (2015). 
	% and this one is FM doped TI 
		\bibitem{NatMat13_699} Y. Fan \textit{et al}, Nature Mater. \textbf{13}, 699 (2014).  	

% this is like the spiritual origin of the Kubo in this paper 
	\bibitem{PRB89_165307} A. Sakai and H. Kohno, Phys. Rev. B \textbf{89}, 165307 (2014). 


% Debashree and co's thin film paper 
	\bibitem{JPD49_135003} A. Menon, D. Chowdhury and B. Basu, J. Phys. D : Appl. Phys. \textbf{49}, 135003 (2016). 
% TI thin film papers
	\bibitem{PRB80_205401}  J. Linder, T. Yokoyama and Al Sodb\o,  Phys. Rev. B \textbf{80}, 205401 (2009). 
	\bibitem{PRB81_041307} C.-X. Liu \textit{et al}, Phys. Rev. B \textbf{81}, 041307 (2010). % SC Zhang kahkia 
	\bibitem{PRB81_115407} H.-Z. Lu \textit{et al}, Phys. Rev. B \textbf{81}, 115407 (2010).  % Shen SQ kahkia 


% that TI thin film got  QPT
	\bibitem{PRB83_195413} A. A. Zyuzin and A. A. Burkov, Phys. Rev. B \textbf{83}, 195413 (2011). 	


% my no disorder TI thin film paper 
	\bibitem{Ar1606_03812} Z. B. Siu, M. B. A. Jalil and S. G. Tan, (unpublished),  arXiv:1606.03812 (2016). 

	\bibitem{Manchon} D. Fang \textit{et al}, Nat. Nanotech. \textbf{6}, 413 (2011); L. Hang \textit{et al}, Phys. Rev. B \textbf{81}, 134402 (2015); L. Hang, X. Wang and A. Manchon, Phys. Rev. B \textbf{93}, 035417 (2016). 


%  my TI thin film with hex warping paper 
	\bibitem{AIPAdv6_055706} Z. B. Siu, S. G. Tan and M. B. A. Jalil, AIP Adv. \textbf{6}, 055706 (2016). 	
	\bibitem{PRB82_045122} C.-X. Liu \textit{et al}, Phys. Rev. B \textbf{82}, 045122 (2010). % model Hamiltonian
	
	
		
		
% our groups B x db B paper
\bibitem{Group} T. Fujita, M.B.A. Jalil and S.G. Tan, J. Phys. Soc. Jpn. \textbf{78}, 104714 (2009), T. Fujita, M.B.A. Jalil and S.G. Tan, New J. Phys. \textbf{12}, 013016 (2010), T. Fujita \textit{et al}, J. Appl. Phys. \textbf{110}, 121301 (2011). , S.G. Tan and M.B.A. Jalil, J. Phys. Soc. Jpn. \textbf{82}, 094714 (2013), S.G. Tan \textit{et al}, Sci. Rep. \textbf{5}, 18409 (2015).		
% Kurebayashi
\bibitem{NatNano9_211} H. Kurebayashi \textit{et al}, Nat. Nanotechnology \textbf{9}, 211 (2014). 

\end{thebibliography}
\end{document}